\documentclass[letter,notitlepage,12pt,showkeys]{revtex4-1}

\usepackage[]{graphicx}
\usepackage{amsmath}
\usepackage{cmap}

\usepackage{longtable}
\usepackage{ltxtable}
\usepackage{tabularx}
\usepackage{array}

\newcommand{\lnA}{\left<\ln{A}\right>}
\newcommand{\ethr}{\epsilon_{\text{thr.}}}

\begin{document}

\title{Cosmic ray spectrum in the energy range $10^{15}-10^{18}$~eV and the second knee according to the small Cherenkov setup at the Yakutsk EAS array}

\author{S.~P.~Knurenko}
\author{Z.~E.~Petrov}
\author{R.~Sidorov}
\author{I.~Ye.~Sleptsov}
\author{S.~K.~Starostin}
\author{G.~G.~Struchkov}
\affiliation{Yu.G. Shafer Institute of cosmophysical research and aeronomy SB RAS}
\email{s.p.knurenko@ikfia.ysn.ru}

\begin{abstract}
  From the data on Cherenkov light from EASs collected over the period of more than 15 years, the spectrum of cosmic rays was obtained in the energy range $10^{15}-10^{18}$~eV. This spectrum has two features: at $\sim 3 \times 10^{15}$~eV (first knee) and at $\sim 10^{17}$~eV (second knee). The first knee is characterized by the indexes $\gamma_1 = 2.7 \pm 0.03$ and $\gamma_2 = 3.12 \pm 0.03$ and the second knee~--- by $\gamma_1 = 2.92 \pm 0.03$ and $\gamma_2 = 3.24 \pm 0.04$. In the second case the difference amounts to $\Delta\gamma_{23} = 0.32 \pm 0.03 \pm 0.05$ which is less than the case of the first knee $\Delta\gamma_{12} = 0.42 \pm 0.03 \pm 0.05$ (here dual errors represent statistics and systematics correspondingly). The lesser difference $\Delta\gamma_{23}$ can be explained with the influx of cosmic rays from Meta-galaxy and, hence, with some increase of the cosmic rays intensity in the energy range $5 \times 10^{16}-10^{18}$~eV, which compensates the escape of heavier nuclei from the Galaxy. The presence of the second knee could be confirmed by abrupt change in mass composition from from $\lnA \sim 3$ at $\sim 10^{17}$~eV to $\lnA \sim 1.5$ at $10^{18}$~eV found from the analysis of longitudinal development of EAS.
\end{abstract}


\maketitle

\section{Introduction}

Determination of the exact shape of the cosmic ray (CR) spectrum within the energy range $10^{16} - 10^{18}$~eV still remains one of the problems of current interest in modern CR physics. This fact is mostly connected to the small aperture of compact arrays in energy range $10^{16} - 10^{18}$~eV and also to impossibility of measuring the spectrum in this range with giant arrays. On the other hand, this region if of the utmost interest concerning the presence of irregularities arising from generation of thin structure in the spectrum at $\sim 5\times 10^{15} - 10^{17}$~eV~\cite{bib1}. The origin and sources of CR of these energies are yet to be identified.

To solve this problem one needs to measure the power spectrum of extensive air showers (EAS) with a good precision (or other equivalents of the power). It is also necessary to determine the function connecting measured parameter of power with energy of primary particle $E_{0}$~\cite{bib2, bib3}. Being an array that can bridge the gap between compact and giant arrays, Yakutsk EAS array fits the task~\cite{bib4, bib5}; besides, it measures all main shower components (electrons, muons and cherenkov light).

\section{Methodical issues and shower selection}

Yakutsk EAS array measures shower particles with the use of scintillation or Cherenkov light detectors separated by sensible distance. In the case of the small Cherenkov setup it is $10-400$~m, for the main array it is $100-2000$~m from the shower axis. In this case it is possible to obtain samples of charged particle density $\rho_{s}(300)$ and $\rho_{s}(600)$ and Cherenkov light flux $Q(150)$ and $Q(400)$. According to simulation of measurements at Yakutsk array~\cite{bib6}, these EAS characteristics have minimal fluctuations and can be utilized as classification parameters during showers selection for obtaining the spectrum. Also, since they are proportional to he primary energy $E_{0}$ of a shower, they may serve as power equivalent.

Nevertheless, reconstructing the EAS spectrum from these parameters is a difficult task, since: 1)\,showers are selected by two different triggers and this affects the resulted spectrum; 2)\,precision of measurements and fluctuations in shower development must be taken into account when calculating the effective area ($S_{\text{eff.}}$) within the fixed range of the classification parameter; 3)\,reduction of shower parameters to vertical direction requires the exact value of charged particles attenuation length; 4)\,when selecting or processing events by Cherenkov light emission, one needs to know parameters of the atmosphere.

\begin{figure}
  \centering
  \includegraphics[width=0.85\textwidth, clip]{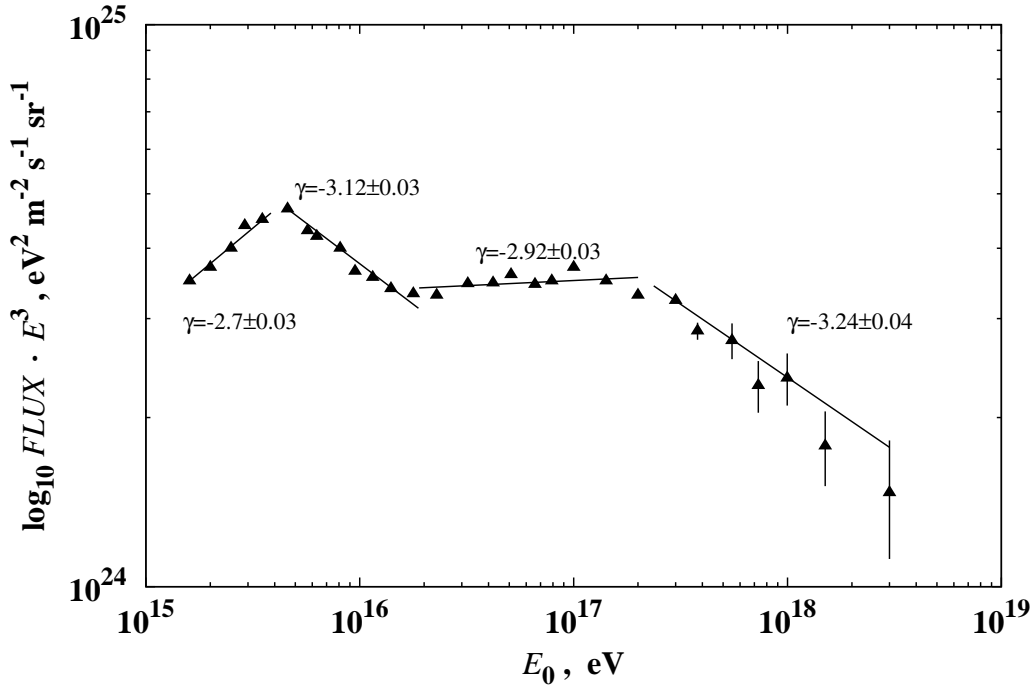}
  \caption{Cosmic ray spectrum in the region $10^{16}-10^{18}$~eV derived from Yakutsk data.}
  \label{fig1}
\end{figure}

Estimated precision of measurements obtained in simulation~\cite{bib6} together with lateral distributions of charged component, muons ($\ethr \ge 1$~GeV) and Chrenkov light derived from significant statistics~\cite{bib7} allowed the Yakutsk array to solve the methodical issues mentioned above with a better precision, then in earlier works on the EAS spectrum~\cite{bib8}.

The data for the spectrum were selected by several triggers. In the energy range $10^{15} - 10^{16}$~eV~--- it was stations with $50$ and $100$~m spacing and in $10^{16} - 10^{18}$~eV~--- stations with $250$ and $500$~v spacing. Transitional effect between two triggers was taken into account during the spectrum reconstruction via Monte Carlo simulation of events selection, which has shown that it amounts to $\sim 20$\,\%. Also this work utilizes a new data set for the energy $10^{17} - 10^{18}$~eV obtained by the small Cherenkov setup during recent years. As a result, the power spectrum was obtained with semi-equal precisions of estimated $Q(150)$ parameter within the wide range of its values (see Fig.~\ref{fig1}). It was transformed into energy spectrum with the use of balancing the shower energy~\cite{bib3}. Numerical values are listed in the Table~\ref{tab1}.
\begin{longtable}{p{0.23\textwidth}p{0.32\textwidth}l}
  \caption{Numerical values for CR spectrum (see Fig.~\ref{fig1}).}
  \label{tab1} \\
    \hline
    $E_{0}$, eV & $\log_{10}{F} \cdot E_0^3$, eV$^2$ m$^{-2}$ sr$^{-1}$ & error \\
    \hline
    \endfirsthead
    %\hline
    %\multicolumn{3}{l}{\sl cont.} \\
    \hline
    $E_{0}$, eV & $\log_{10}{F} \cdot E_0^3$, eV$^2$ m$^{-2}$ sr$^{-1}$ & error \\
    \hline
    \endhead
    \hline
    \multicolumn{3}{l}{{\sl continued on next page}} \\
    \endfoot
    \endlastfoot
    $1.60 \times 10^{15}$ & $3.50 \times 10^{24}$ & -- \\
    $2.00 \times 10^{15}$ & $3.70 \times 10^{24}$ & -- \\
    $2.50 \times 10^{15}$ & $4.00 \times 10^{24}$ & -- \\
    $2.90 \times 10^{15}$ & $4.39 \times 10^{24}$ & -- \\
    $3.50 \times 10^{15}$ & $4.50 \times 10^{24}$ & -- \\
    $4.60 \times 10^{15}$ & $4.70 \times 10^{24}$ & -- \\
    $5.70 \times 10^{15}$ & $4.30 \times 10^{24}$ & -- \\
    $6.30 \times 10^{15}$ & $4.20 \times 10^{24}$ & -- \\
    $8.10 \times 10^{15}$ & $4.00 \times 10^{24}$ & -- \\
    $9.50 \times 10^{15}$ & $3.64 \times 10^{24}$ & -- \\
    $1.15 \times 10^{16}$ & $3.55 \times 10^{24}$ & -- \\
    $1.40 \times 10^{16}$ & $3.39 \times 10^{24}$ & -- \\
    $1.78 \times 10^{16}$ & $3.32 \times 10^{24}$ & -- \\
    $2.29 \times 10^{16}$ & $3.30 \times 10^{24}$ & -- \\
    $3.20 \times 10^{16}$ & $3.46 \times 10^{24}$ & -- \\
    $4.20 \times 10^{16}$ & $3.47 \times 10^{24}$ & -- \\
    $5.10 \times 10^{16}$ & $3.59 \times 10^{24}$ & -- \\
    $6.60 \times 10^{16}$ & $3.45 \times 10^{24}$ & -- \\
    $7.90 \times 10^{16}$ & $3.50 \times 10^{24}$ & -- \\
    $1.00 \times 10^{17}$ & $3.70 \times 10^{24}$ & -- \\
    $1.42 \times 10^{17}$ & $3.50 \times 10^{24}$ & -- \\
    $2.00 \times 10^{17}$ & $3.30 \times 10^{24}$ & -- \\
    $3.00 \times 10^{17}$ & $3.23 \times 10^{24}$ & -- \\
    $3.80 \times 10^{17}$ & $2.85 \times 10^{24}$ & $1.00 \times 10^{23}$ \\
    $5.50 \times 10^{17}$ & $2.74 \times 10^{24}$ & $2.00 \times 10^{23}$ \\
    $7.30 \times 10^{17}$ & $2.28 \times 10^{24}$ & $2.40 \times 10^{23}$ \\
    $9.98 \times 10^{17}$ & $2.35 \times 10^{24}$ & $2.50 \times 10^{23}$ \\
    $1.50 \times 10^{18}$ & $1.78 \times 10^{24}$ & $2.70 \times 10^{23}$ \\
    $3.00 \times 10^{18}$ & $1.47 \times 10^{24}$ & $3.50 \times 10^{23}$ \\
    $3.80 \times 10^{17}$ & $2.85 \times 10^{24}$ & $1.00 \times 10^{23}$ \\
    $5.50 \times 10^{17}$ & $2.74 \times 10^{24}$ & $2.00 \times 10^{23}$ \\
    $7.30 \times 10^{17}$ & $2.28 \times 10^{24}$ & $2.40 \times 10^{23}$ \\
    $9.98 \times 10^{17}$ & $2.35 \times 10^{24}$ & $2.50 \times 10^{23}$ \\
    $1.50 \times 10^{18}$ & $1.78 \times 10^{24}$ & $2.70 \times 10^{23}$ \\
    $3.00 \times 10^{18}$ & $1.47 \times 10^{24}$ & $3.50 \times 10^{23}$ \\
    \hline
\end{longtable}

\section{Spectrum shape according to the Small Cherenkov Setup data}

The spectrum obtained by the small setup has two peculiarities (see Fig.~\ref{fig1}): at $\sim 3 \times 10^{15}$~eV (first knee) and at $\sim 10^{17}$~eV (second knee). The first knee is characterized by indexes $\gamma_1 = 2.70 \pm 0.03$ and $\gamma_2 = 3.12 \pm 0.03$, the second knee~--- by indexes $\gamma_1 = 2.92 \pm 0.03$ and $\gamma_2 = 3.24 \pm 0.04$. For the second case, the difference amounts to $\Delta\gamma_{23} = 0.32 \pm 0.03 \pm 0.05$ which is less than at the first knee: $\Delta\gamma_{12} = 0.42 \pm 0.03 \pm 0.05$. The absence of a sharp kink in the spectrum, like one at the first knee, is probably connected to the presence of a component of extragalactic origin. For example, a lesser difference $\Delta\gamma_{12}$ could be explained with the influx of CR from meta-galaxy (Berezinsky et al, 2004)~\cite{bib9} and, hence, slightly increased intensity of CR at $5\times 10^{16} - 3 \times 10^{18}$~eV. This increase is compensated by abrupt drop of intensity associated with escape of heavier nuclei from the galaxy. In this case it's possible to state, that transition between galactic and meta-galactic CR most probably lies in energy interval $3 \times 10^{17} - 5 \times 10^{18}$~eV.

\begin{figure}
  \centering
  \includegraphics[width=0.85\textwidth, clip]{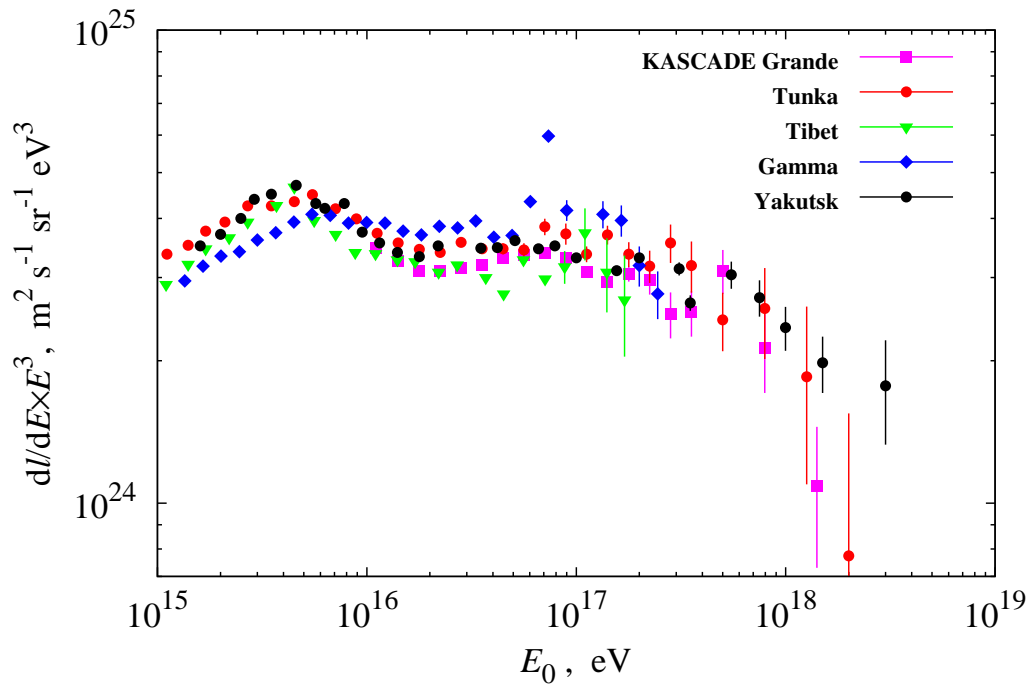}
  \caption{CR spectrum summary. Squares~--- KASCADE Grande, red circles~--- Tunka, blue diamonds~--- Gamma, green triangles~--- Tibet~III, black circles~--- Yakutsk.}
  \label{fig2}
\end{figure}

\begin{figure}
  \centering
  \includegraphics[width=0.85\textwidth, clip]{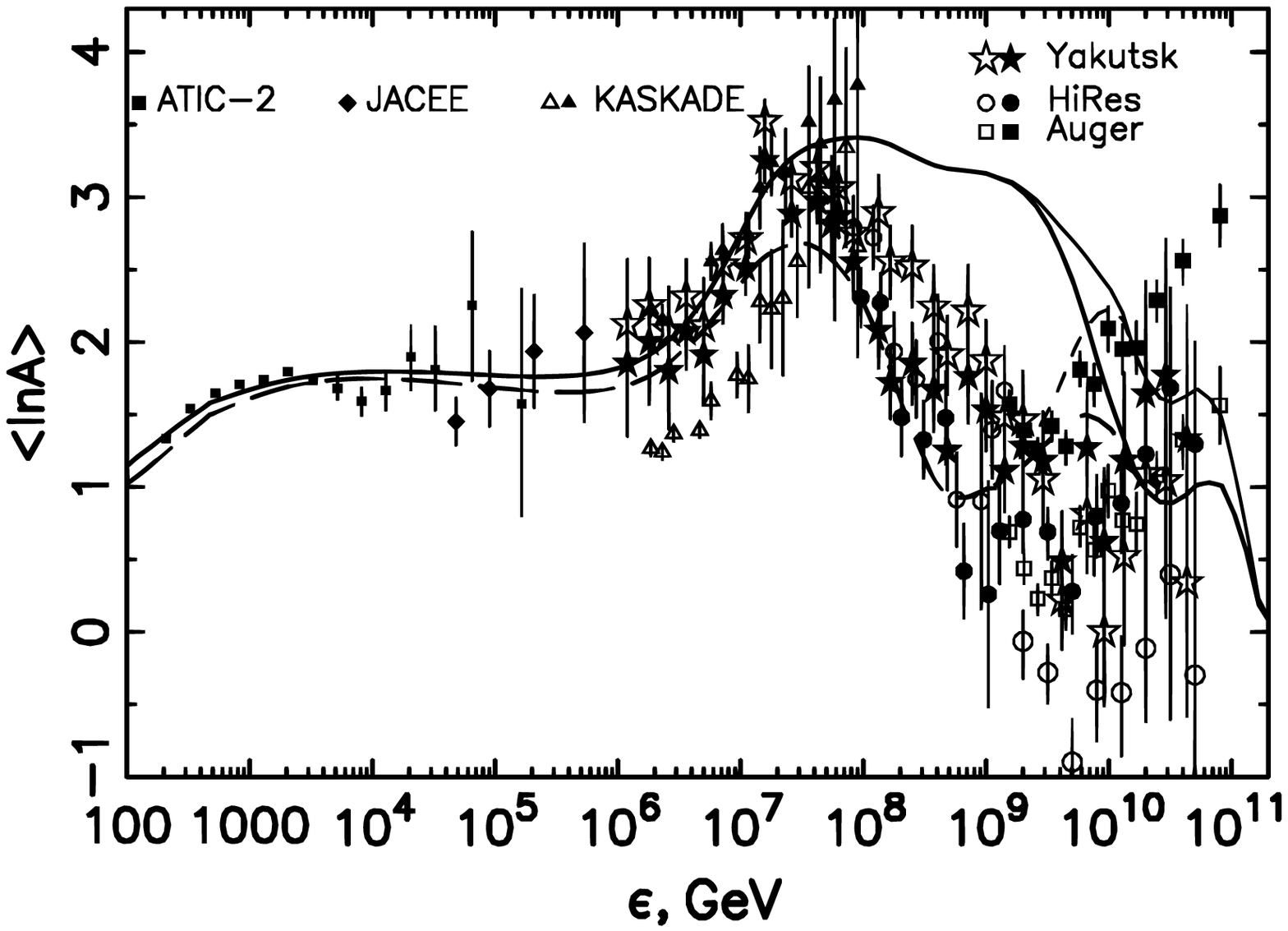}
  \caption{Mass composition obtained by various experiments within the frameworks of hadron interaction models QGSJETII-03 and SIBYLL-2.1. Yakutsk data have been obtained from measurement of Cherenkov light emission by the Small Cherenkov Setup and main array.}
  \label{fig3}
\end{figure}

Summary on resulting spectrum is shown on Fig~\ref{fig2}. It reproduces well the spectrum shape obtained in Yakutsk experiment and does not contradict the second knee hypothesis stated above. And the abrupt change of CR mass composition from $\lnA \sim 3$ at $\sim 10^{17}$~eV to $\lnA \sim 1.5$ $\sim 10^{18}$~eV (see also Fig.~\ref{fig3}) might prove the presence of the second knee in the spectrum.

\section{Results and discussion}

There are several models explaining complex shape of the spectrum. In the sixties of the past century the existence of cycles in generation of the CR spectrum was proposed by Peters. These cycles are connected with the influence of magnetic rigidity on nuclei with different charges, the so-called Peters cycle~\cite{bib1}. It is known from measurements on compact arrays that the spectrum continues after the first kink but with lesser intensity and rigidity. It allowed to suggest the presence of a new powerful sources and, hence, new unknown component of cosmic rays and the second kink in the spectrum at $\sim 8 \times 10^{16}$~eV between ankle and knee. This component was called ``component B'' by Hillas~\cite{bib11}.

In the interpretation of the spectrum developed by Nagano and Watson~\cite{bib12} the shape of the spectrum around $10^{16} - 10^{18}$~eV was explained by presence of the ``component B'' in the galaxy. And at energy above $10^{18}$~eV including the ankle it was proposed that the spectrum is generated by meta-galactic cosmic rays.

According to the model proposed by Berezinsky et al, spectrum shape in energy region above $10^{17}$~eV might be influenced by meta-galactic CR accelerated in sources up to $10^{20}$~eV and dominated by proton component. In this case there is no need for the ``component B'' for explanation of the spectrum shape in the region of moderate energies. As a result the transitional border lies between $10^{16}$~eV and $10^{18}$~eV~\cite{bib13}.

The summary spectrum constructed from the data from various compact arrays (see Fig.~\ref{fig2}) demonstrates abrupt decrease of CR intensity after the energy $\sim 10^{17}$~eV. Most probably it is the second knee. The difference in the spectrum index after the kink $\Delta\gamma_{12} = 0.32$ and confirms our conclusion.

It is possible to explain the continuation of the spectrum and small number of particles after the second knee with transitional border between galactic and extra-galactic CR. The analysis of the mass composition (see Fig.~\ref{fig3}) also confirms this. According to Fig.~\ref{fig3}, the mass composition also abruptly changes from heavier to lighter at $\sim 10^{17}$~eV and stays light up to energies $\sim 10^{19}$~eV. According to calculations performed by Berezhko~\cite{bib10} such mass composition and its energy dependency in the $10^{16} -10^{18}$~eV region can be explained with the presents of supernova remnants in the galaxy with condition of non-linear CR acceleration on shock fronts and subsequent re-acceleration during their propagation in interstellar medium. CR of higher energies most probably have extra-galactic origin.

\acknowledgments
This work is supported by The Russian Foundation for Basic Research (grant project 12-02-31442\,mol\_a) and by Ministry of Education and Science of the Russian Federation (contract 8404).

\end{document}